# Acoustic methods for measuring bullet velocity

Michael Courtney
Ballistics Testing Group, P.O. Box 24, West Point, NY 10996
*Michael_Courtney@alum.mit.edu*

**Abstract:** This article describes two acoustic methods to measure bullet velocity with an accuracy of 1% or better. In one method, a microphone is placed within 0.1 m of the gun muzzle and a bullet is fired at a steel target 45 m away. The bullet's flight time is the recorded time between the muzzle blast and sound of hitting the target minus the time for the sound to return from the target to the microphone. In the other method, the microphone is placed equidistant from both the gun muzzle and the steel target 91 m away. The time of flight is the recorded time between the muzzle blast and the sound of the bullet hitting the target. In both cases, the average bullet velocity is simply the flight distance divided by the flight time.

## 1. Introduction

Early measurements of bullet velocity relied on the ballistic pendulum which remains an important educational tool for demonstrating conservation of momentum and energy, as well as providing a method for measuring the velocity generated by potato cannons [1]. As firearms design progressed, higher velocities and longer range applications mandated greater accuracy to predict long range trajectories. Thus the optical chronograph was developed which optically detects the bullet passing over fixed points and uses the time interval to compute velocity.

Optical chronographs are simple and affordable, but there are drawbacks. The instrument is placed in the line of fire. Consequently, chronographs are often damaged by bullets. Poor lighting conditions can produce inconsistent chronograph results. Cables and sky screens require time to set up, and putting instruments in front of the firing line is unwelcome at many range facilities.

Soundcards are inexpensive, readily available and included in most personal computers. Digital audio recorders that permit downloading of the waveform to a computer are also common.

## 2. Acoustic methods

The first acoustic method employs a microphone placed within 0.1 m of the muzzle, preferably to the side and slightly behind the muzzle to prevent the microphone from being hit directly by the muzzle blast or by the recoiling firearm. A steel target making a loud noise when hit is placed 45.74 m (50 yds) away. Targets mounted to freely resonate work better than targets whose vibration is quickly damped.

The audio digitizer should be operated with software (such as Audacity [2]) that allows viewing the waveform to determine the time difference between the beginning of the muzzle blast and the target sound. A Vernier LabPro with a microphone works well also. Under some conditions, the target sound is masked by reverberation of the muzzle blast. Increasing the target distance gives the muzzle blast more time to dissipate before the target sound arrives back at the microphone. If the sound of the target hit is masked by noise, the time of impact can be determined by making a spectrogram of the sound waveform.

The bullet time of flight ($t_{bullet}$) is determined by subtracting the time it takes sound to travel from the target back to the microphone ($t_{sound}$) from the time recorded between the muzzle blast and target sound ($t_{total}$),

$$t_{bullet} = t_{total} - t_{sound}. \quad (1)$$

The time for sound to travel back from the target to the microphone is the distance ($d$) divided by the velocity of sound ($c_{sound}$),

$$t_{sound} = d / c_{sound}. \quad (2)$$

$c_{sound}$ is roughly 331 m/s at 0°C in a neutral atmosphere. It varies with temperature approximately according to

$$c_{sound} = 331 \text{ m/s} + 0.6 \text{ (m/s°C) } T, \quad (3)$$

where $T$ is the temperature in Celsius. In addition to having an accurate distance measurement (within 0.1 m), one needs the ambient temperature within 1°C. If the bullet path is at least 0.5 m above the ground, ground effects on $c_{sound}$ can be ignored without contributing more than 1% to the error in determination of bullet velocity.

Once the time of flight of the bullet is determined, the average bullet velocity is

$$V_{bullet} = d / t_{bullet}. \quad (4)$$

The greatest source of error in this method is the uncertainty in the speed of sound under field conditions. The second method places the microphone equidistant from target and muzzle. Thus, the sound travel times from muzzle to microphone and from target to microphone are equal and the flight time is simply the recorded time between the muzzle blast and the target strike [3].

## 3. Results and Analysis

The load tested with the microphone adjacent to the muzzle is a .22 Long Rifle (.22 LR). The software begins recording when a loud noise is detected, so the beginning of the muzzle blast is at $t = 0$. (The microphone signal goes from its low noise value to saturation in a single sample.) Figure 1 shows several prominent peaks corresponding to muzzle blast and echoes (probably from nearby posts and trees). The peak at $t = 0.2594$ s records the target strike.

The target distance is 45.74 m (50 yds). The speed of sound at 24.4°C is 345.6 m/s. Consequently, the time for the sound to return from target to microphone is $t_{sound} = 0.1323$ s.

Zooming in on the waveform shows the time between the muzzle blast and target sound is $t_{total} = 0.2594$ s. Subtracting the time for the sound to return to the microphone gives $t_{bullet} = 0.1271$ s. Substituting into equation (4) yields $V_{bullet} = 359.9$ m/s.

For comparison, the velocity measured on an optical chronograph is 387.1 *m/s*. These measurements do not seem to agree, because the acoustic method measures the average velocity over the flight which is slightly lower than the muzzle velocity due to air resistance slowing the bullet in flight.

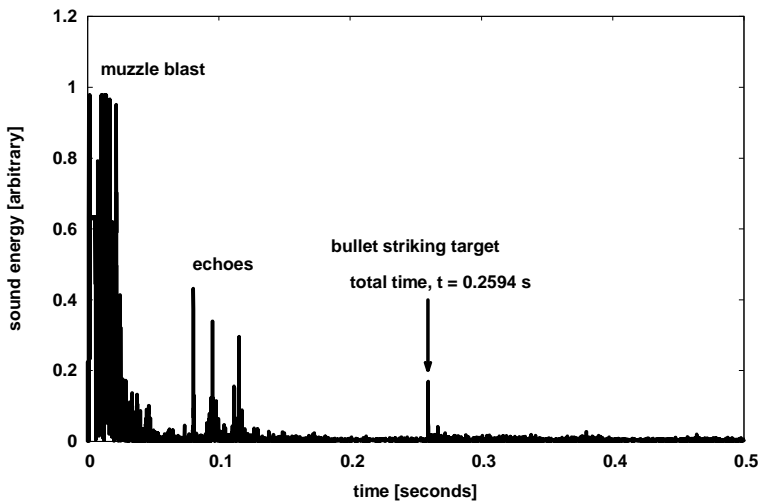

*Figure 1: Sound energy vs. time for .22 LR shot with microphone placed adjacent to gun muzzle. Sound energy is proportional to the square of the microphone voltage.*

The ballistic coefficient is a dimensionless quantity describing how air resistance slows a bullet in flight. Ballistic calculators are available to compute time of flight and trajectory for a given muzzle velocity, ballistic coefficient, and ambient conditions [4]. The ballistic coefficient roughly corresponds to the fraction of 915 m (1000 yds) that a bullet must travel to lose 50% of its initial kinetic energy. For example, a bullet with a ballistic coefficient of 0.4 loses roughly 50% of its energy at a range of 366 m (400 yds).

Together with the muzzle velocity and ambient conditions, this bullet's ballistic coefficient of *0.11* gives a predicted time of flight of *0.1264 s* for *45.74 m*, thus an average velocity of *361.9 m/s.* Consequently, the acoustic measurement of bullet average velocity is in excellent agreement *(0.55%)* with the value determined using the chronograph to predict the time of flight.

*Table 1: Comparison of average bullet velocities determined by acoustic method (microphone adjacent to gun) and by using muzzle velocity and ballistic coefficient.*

| Shot | $V_{bullet}^{acoustic}$ | $V_{bullet}^{chronograph}$ | Difference |
|---|---|---|---|
| 1 | 359.9 m/s | 361.9 m/s | -0.55% |
| 2 | 360.2 m/s | 361.3 m/s | -0.30% |
| 3 | 358.7 m/s | 359.0 m/s | -0.08% |
| 4 | 357.3 m/s | 357.9 m/s | -0.16% |

Repeating the above procedure for three additional test shots produces the acoustic and chronograph average velocities shown in Table 1. (The chronograph average velocity is determined by using the ballistic coefficient, muzzle velocity, and ambient conditions to compute the time of flight.) The average velocity determined from the acoustic method is consistently within 1% of the average velocity determined with the chronograph. This procedure yields comparable accuracy when used with much faster bullets (> 800 m/s).

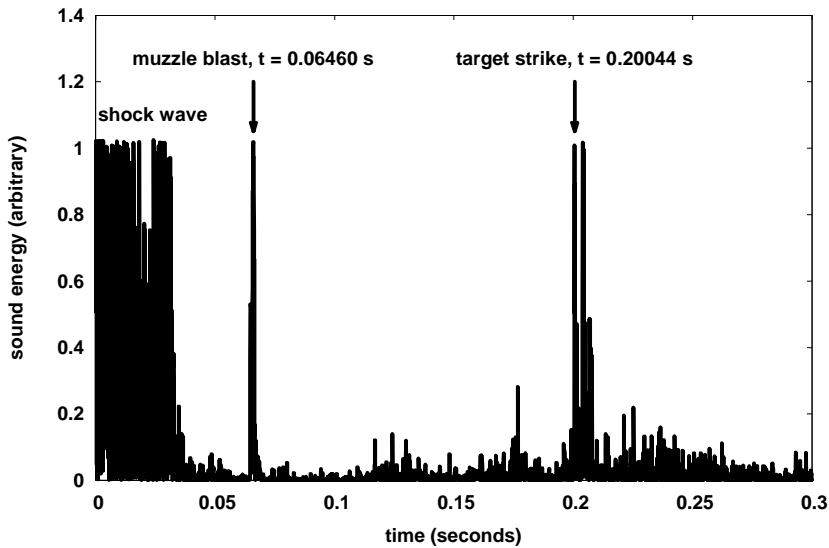

*Figure 2: Sound energy vs. time for .223 Remington rifle shot with microphone placed equidistant from gun muzzle and steel target. Sound energy is proportional to the square of the microphone voltage.*

The equidistant method was tested with a mild .223 Remington load. The target distance was 91.48 m (100 yards). Figure 2 shows that the shockwave reaches the microphone first, triggers the sampling, and briefly saturates the signal.

The muzzle blast is recorded at *0.06460 s* followed by the target strike at *0.20044 s.* The resulting time of flight is the difference, $t_{bullet}$ = *0.13584 s*, which yields an average velocity of $V_{bullet}$ = *673.4 m/s.* The measured muzzle velocity is 731.5 m/s which (together with the ballistic coefficient of 0.218 and ambient conditions) predicts an average bullet velocity of 687.6 m/s over the flight distance. Consequently, the acoustic measurement is in excellent agreement (0.77%) with the value of average velocity predicted with the chronograph.

Repeating the above procedure for three additional test shots produces the acoustic and chronograph average velocities shown in Table 2, which shows that the average velocity determined from the acoustic method with the equidistant microphone is consistently within 1% of the average velocity determined with the chronograph.

Table 2: Comparison of average bullet velocities determined by acoustic method (equidistant microphone) and by using muzzle velocity and ballistic coefficient.

| Shot | $V_{bullet}^{acoustic}$ | $V_{bullet}^{chronograph}$ | Difference |
|---|---|---|---|
| 1 | 673.4 m/s | 678.6 m/s | -0.77% |
| 2 | 676.6 m/s | 673.6 m/s | 0.44% |
| 3 | 673.8 m/s | 678.6 m/s | -0.71% |
| 4 | 698.5 m/s | 701.5 m/s | -0.43% |

**4. Discussion**

The equidistant method should be more accurate than placing the microphone adjacent to the gun because uncertainties in the speed of sound under ambient conditions are no longer a factor. However, in comparing the chronograph results with the equidistant acoustic method, the uncertainties from using the chronograph, ballistic coefficient, and ambient conditions to predict average bullet velocity are probably larger than the errors inherent in the equidistant technique, so a much more accurate method (radar) is needed to determine the lower error bound of the equidistant method. Since most loads show shot-to-shot velocity variations of 1% or more, the level of accuracy already confirmed should be sufficient for most applications, including determination of ballistic coefficients when coupled with a chronograph [3].

However, the equidistant method has the drawback of putting equipment in front of the firing line, creating a risk of damage from inaccurate projectiles (shotgun pellets, some pistol bullets) and secondary projectiles (sabots and shotgun wads) that separate from the main projectile(s) as well as not being allowed at some shooting ranges. For supersonic projectiles with velocities below 400 m/s (most shotguns, many pistols, some rifles), placing the microphone equidistant from the muzzle and target also creates difficulty distinguishing the muzzle blast from the passing shock wave because they arrive at nearly the same time.

The method with the microphone adjacent to the muzzle has potential commercial application using a soundcard and automating computations in software. However, reliable commercial application requires discriminating the target sound in cases where the muzzle blast and other sounds prevent the clear discrimination of a peak when the bullet hits the target. This can be accomplished with Fourier analysis where the program learns the spectral signature of the target and detects the time at which this spectral signature occurs in the sound waveform. Alternatively, a high-pass filter would eliminate low-frequency components of the boom and be a simpler approach to enhancing ease of discriminating the sound of the target hit.

Two methods for acoustic determination of average bullet velocity have been presented. Placing the microphone adjacent to the muzzle is more generally applicable. Both methods provide reasonable accuracy, but there are drawbacks. Loud noise can interfere with the measurement, which necessitates more detailed spectral analysis or filtering.

If the ballistic coefficient, range, and ambient conditions are known, determining any one of: (i) bullet muzzle velocity, (ii) average bullet velocity, or (iii) terminal bullet velocity allows the other two to be determined also. Bullet trajectory, wind drift, and time of flight are determined by average velocity. Average velocity also dominates acoustic reconstructions of shooting events. In contrast, terminal ballistics depends on the terminal velocity, which determines bullet penetration, expansion, and terminal energy. Terminal velocity can also aid determining range in shooting event reconstructions based on penetration or wound characteristics.

Finally, a two dimensional generalization of these ideas using triangulation has possible applications in the acoustic reconstruction of criminal shooting events. The sound of the muzzle blast has long been used to determine possible locations of the shooter via triangulation of the muzzle blast received by multiple microphones. Using the sound of the target strike allows determination of unknown factors in cases where audio recordings are available and the microphone locations are known. If the locations of muzzle, target, and microphone are known, an unknown muzzle velocity can be determined, thus narrowing the possibilities for the type of firearm. If the muzzle velocity is known from recovered cartridge cases or other evidence, it is possible to determine the distance from the shooter to the target.

**About the Author**

*Michael Courtney* earned a PhD in experimental Physics from the Massachusetts Institute of Technology. He has served as the Director of the Forensic Science Program at Western Carolina University as well as a Physics Professor, teaching college level Physics, Statistics, and Forensic Science. He has extensive experience in analysis of complex systems, instrumentation and measurement, and data analysis. Michael founded the Ballistics Testing Group in 2001 to study incapacitation ballistics and the reconstruction of shooting events. See: www.ballisticstestinggroup.org